\documentclass[aps,twocolumn, prb,groupedaddress,showkeys]{revtex4}
\usepackage{amssymb}
\usepackage{amsmath}
\usepackage{amsmath}
\usepackage{amsmath,amssymb,bm,graphicx}

\begin{document}

\title{ Double-slit and electromagnetic models to complete quantum mechanics}
\author{Jayme De Luca}
\email{deluca@df.ufscar.br}
\affiliation{Universidade Federal de S\~{a}o Carlos, \\
Departamento de F\'{\i}sica\\
Rodovia Washington Luis, km 235\\
Caixa Postal 676, S\~{a}o Carlos, S\~{a}o Paulo 13565-905}
\date{\today }

\begin{abstract}
We analyze a realistic microscopic model for electronic scattering based on
the neutral differential delay equations for point charges of the
Wheeler-Feynman electrodynamics. We propose a microscopic model according to
the electrodynamics of point charges, complex enough to describe the
essential physics. Our microscopic model reaches a simple qualitative
agreement with the experimental results as regards interference in
double-slit scattering and in electronic scattering by crystals. We discuss
our model in the light of existing experimental results, including a
qualitative disagreement found for the double-slit experiment. We discuss an
approximate solution for the neutral differential delay equations of our
model using piecewise-defined (discontinuous) velocities for all charges and
piecewise-constant-velocities for the scattered charge. Our approximation
predicts the De Broglie wavelength as an inverse function of the incoming
velocity and in the correct order of magnitude. We explain the scattering by
crystals in the light of the same simplified modeling with Einstein-local
interactions. We include a discussion of the qualitative properties of the
neutral differential delay equations and the boundary-value variational method 
of electrodynamics to stimulate future
experimental tests on the possibility to complete quantum mechanics with
electromagnetic models.
\end{abstract}

\maketitle



\section{\protect\bigskip Introduction}

We discuss an attempt to complete quantum mechanics (QM) with an
Einstein-local deterministic theory along with an analysis of realistic
models for electronic diffraction. We introduce time-dependent interactions
with the microscopic electronic trajectories inside the measuring
apparatuses as a five-body-problem. We find that a microscopic model with
delayed long-range interactions is enough to explain qualitatively the
electronic scattering experiments. We analyze the microscopic models with
the equations of motion for point charges of the Wheeler-Feynman
electrodynamics \cite{Fey-Whe}, whose complex qualitative behavior is worth
stating clearly. The electromagnetic equations of motion are neutral
differential delay equations and have a qualitative behavior quite far from
ordinary differential equations (ODE) \cite{BellenZennaro,JackHale,Hans-Otto}%
. We \ include a brief discussion of some generic qualitative properties of
neutral differential delay equations relevant for the microscopic modeling.
As with any modeling, if one starts with a narrow enough setup it is easy to
obtain impossibility proofs, and the next step is usually harder, i.e., to
define a wider minimum setup. We discuss our models in the light of
available experimental work \cite{Mahalu} and a finite variational method
for the Wheeler-Feynman electrodynamics\cite{JMP2009}.

No single realistic dynamical model based on electrodynamics has been
analyzed down to microscopic detail up to now. The existing gap has two
causes (i) electrodynamics itself was not a finished theory in the early
20th century; the first equation of motion for a point-charge \cite{Dirac}
appeared in 1938 and a form free of self-interaction \cite{Fey-Whe} only in
1945 and (ii) the theory of dynamical systems was not out in the early days
of point-charge-electrodynamics, neither tools existed to understand the
complex dynamics described by differential delay equations. On the other hand there is
the popular quantum schism created by failed naive attempts to complete
quantum mechanics with a Galilei-invariant mechanics based on ordinary
differential equations with globally $C^{\infty }$ trajectories and
instantaneous interactions. Quantum mechanics (QM) was successfully
developed as a Machian probabilistic theory, and there are several hints
that a dynamics with Newtonian ordinary differential equations is \emph{not}
sufficient to complete QM. Even though the simplest quantum operators are
pedagogically constructed by analogy with Galilei-invariant mechanics, a
classical dynamics with ordinary differential equations is useless either to
complete QM \emph{or} to approximate the neutral differential delay
equations of motion of point-charge-electrodynamics, as we show here. It is
known that the no-go-theorems of Bell type fail for non-instantaneous
interactions \cite{Karl} even without invoking the generic properties of
neutral differential delay equations and variational 
methods with boundary conditions in the future,
as discussed here. Nevertheless, there has been no attempt to complete QM with detailed
modeling as regards the double-slit experiment.
\par
This paper is divided as follows: In Section \ref{Section II} we discuss
existing experimental results on double-slit scattering, including a
controlled ballistic experiment that found an unexpected qualitative
disagreement \cite{Mahalu} and the later claims \cite{Jauho} that the
double-slit experiment should not display the simple interference based on a
single De Broglie wave. We give a simplified model based on electromagnetic
trajectories with vanishing far-fields, a model that explains the
qualitative results of interference. In Section \ref{Section III}
we discuss the equations of motion of point charges of the Wheeler-Feynman
electrodynamics and the electromagnetic variational method \cite{JMP2009}.
In Section \ref{Section III} we also work out the details of a model for scattering with 
each slit represented by a heavy neutral atom, a system with little recoil
that is not destroyed upon scattering, unlike a model for bowling. We
introduce a simple approximation for the dynamics with (discontinuous)
piecewise-defined-velocities and piecewise-constant-velocities for the
scattered charge. Our model predicts the qualitative dependence of the De
Broglie wavelength on the incoming electronic velocity and calculates the De
Broglie wavelength in the correct order of magnitude.  In Section \ref{Discussions} 
we put the discussions and conclusion. In appendix A we
discuss the qualitative properties of differential delay- and neutral differential delay
equations with generic initial conditions. These qualitative properties when
included in detailed modeling might offer further possible experimental
tests for electromagnetic modeling and the completion of QM. In appendix B 
we discuss some models for scattering by a periodic crystal. The modeling 
again reaches a qualitative agreement with the experimental results.

\bigskip

\section{Electron scattered by a double-slit}

\label{Section II}

\bigskip

Usually, the quantum explanation of double-slit diffraction is contrasted to
simple billiard-ball-like-one-body-dynamics, and no microscopic model or
time-dependent variables of the apparatus are ever introduced \cite%
{Tonomura,Jonsson}. Let us start by defining this popular no-go modeling of
scattering, henceforth called the billiard-ball-model; This model assumes
the electron interacts with the slit walls by contact forces only. According
to this model, if one closes one slit at a time, an outwardly scattered
electron \emph{must} have gone through the open slit without having
interacted with the closed slit at all (because interaction is by contact
only) \cite{Feynman}. Therefore, the statistical pattern at a far screen
with both slits open should be the sum of two one-slit patterns, in bold
contradiction with the predictions of Schroedinger's equation \emph{and} the
experimental results. On the other hand, the popular quantum model of
double-slit diffraction is the scalar diffraction for the wave equation, a
Schroedinger equation often pedagogically introduced by analogy with
billiard-ball dynamics, with a potential vanishing everywhere but at slit
boundaries. We stress that even though it is time-proven that the analogy
works well from the Galilei-invariant dynamics to abstract quantum
operators, using the analogy backwards to construct a dynamical system to
complete quantum mechanics is far another problem. The reverse analogy \emph{%
from} the scalar wave diffraction \emph{to} the billiard-ball dynamics is
not granted; The difficulties with this \emph{reverse analogy }become
apparent when we consider the\emph{\ realistic }quantum mechanical model for
a \emph{real electron with spin}, approaching a realistic material slit
composed of electrons and protons. 
\par 
The realistic quantum description is as follows;
The incoming electron has a De Broglie distance of influence $\lambda _{DB}$%
,\ so that its wavefunction overlaps the wavefunctions of the electrons at
the material \emph{before} it hits the slit. Because of the exclusion
principle, already at a distance $\lambda _{DB}$, the electronic spin could
flip due to this overlap. Flipping the spin gives the incoming electron an
orbital-angular-momentum-kick of the order of $\hbar $ (electronic spin
carries an angular momentum of $\sqrt{3}\hbar /2$). This change-by-spin-flip
of the orbital-angular-momentum produces diffraction, even \emph{before} the
electrons hits any slit. Notice that quantum mechanics with spin suggests
that the electron suffers an angular-momentum kick without having to pass by
any slit, actually even before going through one of the slits. Therefore,
the careful quantum mechanical analysis of diffraction \emph{with spin }%
suggests a microscopic model with an interaction that can change the orbital
angular momentum at-a-distance (rather than a billiard-ball dynamics). If
the distance between slits is $a$ and the incoming electron has mass $m$ and
velocity $v$, the largest orbital-angular-momentum about an axis
passing by one slit belongs to a trajectory passing by the other slit, i.e., 
$L=amv$ (higher trajectories simply hit the wall and do not pass to the
other side). The largest scattering due to the $\hbar $
angular-momentum-kick of the flipping spin happens when $L=amv\sim \hbar $,
i.e., 
\begin{equation}
a=\frac{\hbar }{mv},  \label{deBroglie}
\end{equation}%
which is the familiar slit-separation for significant scattering. In the
electromagnetic microscopic modeling the electron interacts with both slits
at a distance $\lambda _{DB}$ before passing through one slit, unlike the
interaction by contact of the billiard-ball model. So much to say the
confusion generated by the billiard-ball straw-man-model is unacceptable
moot. 
\par 
The first double-slit experiment \cite{Jonsson} was performed in 1961,
obtaining a scattering pattern in qualitative agreement with the quantum
predictions. The first ballistic experiment with a single electron in the
apparatus at a time \cite{Rodgers} was performed in 1989 by Tonomura et al 
\cite{Tonomura}, obtaining a build-up of fringe patterns in qualitative
agreement with quantum mechanics and the Huygens interference. It seems that
so far no one has claimed quantitative precision beyond order of magnitude
for the double-slit experiment. In 1994 an unexpected periodicity was found
in a controlled ballistic experiment of double-slit scattering \cite{Mahalu}%
. The anomalous periodicity was later explained \cite{Jauho}\ as due to
multiple scattering processes, but the quantum mechanical explanation made
explicit use of a \emph{realistic }model \ \cite{Jauho}. The explanation 
\cite{Jauho} also made evident the need for a realistic quantum modeling and
as a bonus seemed to rule out a precise experimental realization of the
popular idealized experiment. This situation should be contrasted to
electronic diffraction in crystals and periodic structures, where Bragg
directions can be determined very precisely. For the case of a crystal,  
the microscopic modeling captures the essential experimental results even with 
ordinary differential equations, as discussed in appendix B.

\par

In the following we introduce a microscopic model for
double-slit-diffraction consisting of the scattered electron in addition to
an electron bound to each heavy positive charge at each slit site, which are
separated by the slit distance $a$. Our model uses the Wheeler-Feynman
equations of motion of point charges keeping only the far-field
interactions. Our simplification stems from four reasons: (i) we disregard
the short-range Coulomb interactions that are approximately cancelled by
charge neutrality, (ii) we ignore the short-ranged Biot-Savart interactions
that are small at low velocities, (iii) we keep the terms that introduce neutral
delay in the differential equations, i.e., the far-field-couplings
between point charges, which depend on the retarded/advanced accelerations,
and (iv) the far-field interactions have the longest range, representing
naturally the first physical interaction suffered by the electron when
approaching the double-slit and scattering away. We henceforth adopt a
unit system where the speed of light is $c=1$, the electronic charges are $%
e=-1$ and we label quantities related to each electron with indices $k=1,2,3$%
. \ We define trajectories on a Lorentz four-space $\mathcal{L}^{4}$
attached to an inertial frame by Einstein synchronization of clocks. A point
in $\mathcal{L}^{4}$ is defined by a time $t$ and a spatial position $%
\mathbf{x}$ and we parametrize trajectories by time so that each trajectory
is a function $\mathbf{x}_{k}(t_{k})$ $:%
\mathbb{R}
\rightarrow 
\mathbb{R}
^{3}$. As derived in Section \ref{Section III}, the low-velocity equation of motion of
charge $1$ in the far-field-only approximation is 
\begin{eqnarray}
m_{1}\frac{d^{2}}{dt^{2}}\mathbf{x}_{1}(t) &=&\frac{n_{12}^{+}}{2r_{12}^{+}}%
\times \lbrack n_{12}^{+}\times \frac{d^{2}}{dt^{2}}\mathbf{x}%
_{2}(t_{2}^{+})]  \label{motion 1} \\
&&+\frac{n_{12}^{-}}{2r_{12}^{-}}\times \lbrack n_{12}^{-}\times \frac{d^{2}}{%
dt^{2}}\mathbf{x}_{2}(t_{2}^{-})]  \notag \\
&&+\frac{n_{13}^{+}}{2r_{13}^{+}}\times \lbrack n_{13}^{+}\times \frac{d^{2}}{%
dt^{2}}\mathbf{x}_{3}(t_{3}^{+})]  \notag \\
&&+\frac{n_{13}^{-}}{2r_{13}^{-}}\times \lbrack n_{13}^{-}\times \frac{d^{2}}{%
dt^{2}}\mathbf{x}_{3}(t_{3}^{-})].  \notag
\end{eqnarray}%
The equations of motion of the other charges are obtained analogously, by
exchanging indices in Eq. (\ref{motion 1}). \ In Eq. (\ref{motion 1}) unit
vectors $n_{1k}^{\pm }\equiv (\mathbf{x}_{1}(t)-\mathbf{x}_{2}(t_{k}^{\pm
})/|\mathbf{x}_{1}(t)-\mathbf{x}_{k}(t_{k}^{\pm })\mathbf{|}$ point
respectively from the charge's advanced/retarded position $\mathbf{x}%
_{k}(t_{k})$ to the position of charge $1$ at time $t$. In Eq. (\ref{motion
1}) the advanced/retarded times are defined by the implicit formulas 
\begin{equation}
t_{k}^{\pm }=t\pm |\mathbf{x}_{k}(t_{k}^{\pm })-\mathbf{x}(t)\mathbf{|},
\label{deviatime}
\end{equation}%
and the respective distances in lightcone are defined by%
\begin{equation}
r_{1k}^{\pm }=|\mathbf{x}_{k}(t_{k}^{\pm })-\mathbf{x}(t)\mathbf{|}.
\label{distmin}
\end{equation}%
Notice\ that Eq. (\ref{motion 1}) is a \emph{neutral differential delay
equation} with advanced and retarded deviating arguments. In appendix A we
review some qualitative properties of delay- and neutral differential delay
equations and discuss how, for generic initial data, solutions of delay
equations must be defined \emph{piecewise}. Moreover, our approximate
equations of motion depend linearly on the accelerations (see Eq. (\ref%
{motion 1})), so that arbitrary piecewise-constant-velocity solutions can be
added to any balanced solution of the isolated system of bound charges.
In Ref. \cite{unpublished} it is proved that if the bounded motion of two
charges has vanishing far-fields, the dynamics of the bounded charges \emph{must}
have a sewing chain of velocity discontinuities. This gives rise to a
 natural forcing period defined by the sewing chains of the bound trajectories,
as explained in the following.
\par
As discussed in appendix A, in general a neutral differential delay equation
like Eq. (\ref{motion 1}) has piecewise-defined solutions with
discontinuous velocities and accelerations. A discontinuity in the acceleration of particle 
$1$ is propagated in time by Eq. (\ref{motion 1}), as follows; In Figure1 we
illustrate a point $p_{0}$ where the acceleration of charge $3$ is
discontinuous. The equation of motion for charge $1$, Eq. (\ref{motion 1}),
involves the discontinuous past acceleration of charge $3$ on the
right-hand-side, yielding a discontinuous acceleration for charge $1$ at
point $f_{31}(1)$. On the equation of motion for charge $3$ the
discontinuous acceleration of charge $1$ at point $f_{31}(1)$ is on the
right-hand-side and defines another discontinuity for acceleration $3$ at
point $f_{31}(2)$. Successively, this produces a sewing chain of
acceleration discontinuities at times $f_{31}(k)$, as illustrated in Fig. 1,
which define a stroboscopic sampling of each trajectory. The sewing chain
involving trajectories $1$ and $2$ is not illustrated, but it resonates and
is part of the dynamics as well. A controlled ballistic experiment on
double-slit interference found \emph{\ two} wavelengths \cite{Mahalu}; one
near the De Broglie wavelength and another near half the De Broglie
wavelength. Our electromagnetic model with state-dependent-delay naturally
defines \emph{two} wavelengths by the \emph{four} delayed interactions with
the bound electrons, as follows; 
\par
The equation of motion (\ref{motion 1}) and
the respective equations for the other charges have oscillatory solutions
with decreasing wavelength when the electron approaches the inner gate of
scattering. Low-velocity initial conditions approaching the inner gate from
the left-hand-side with an exactly horizontal velocity have equal
advanced/retarded distances to the bound charges, i.e., $r_{12}^{+}\simeq $ $%
r_{12}^{-}\simeq r_{13}^{+}\simeq r_{13}^{-}\equiv r$. For example, along
trajectory $1$ the bouncing times of the advanced interaction are times $%
f_{31}(1)$, $f_{31}(3)$, $f_{31}(5)$ of Fig. 1, which in the far region are
approximately separated by $2r$ and have decreasing time separations until
the limiting period $a$ for a perfectly symmetric horizontal velocity
hitting the center of the inner gate (which never passes to the other side).
In the near region the sewing chain of Fig. 1 starts to resonate also with
the sewing chain between trajectories $1$ and $2$ (not illustrated in Fig.
1), because they share the same frequency. The small vertical velocity
breaks the symmetry in the near region and the state-dependent distances to
the bound charges start to differ, defining which slit the scattered charge
passes through. The sewing chains for a slightly non-horizontal trajectory $%
3 $ excite an effective oscillation with period $L$ superposed to the
dynamics of the interacting bound charges. 
\par
We henceforth assume that the
scattered charge finds a resonant path along which the distances in
lightcone to the bound charges are multiples of each other. The sewing chain
of interactions with the bound charges excite a period $L$ smaller than the
limiting value $a$ of the symmetric centrally horizontal initial condition.
In Section \ref{Section III} we estimate this resonant wavelength $L$ by the distance of
closest approximation between the scattered charge and the atom at the slit
it passes through. The calculations assume the bound charges
are bound to heavy positive charges sitting at each slit end, so that
perturbations imposed by the incoming electron on the bound electrons are
passed to the heavy charges at each site, producing smaller perturbations to
disturb the scattered electron along its outer scattering path.

\bigskip

\begin{figure}[h]
\centering
\includegraphics[scale=0.47]{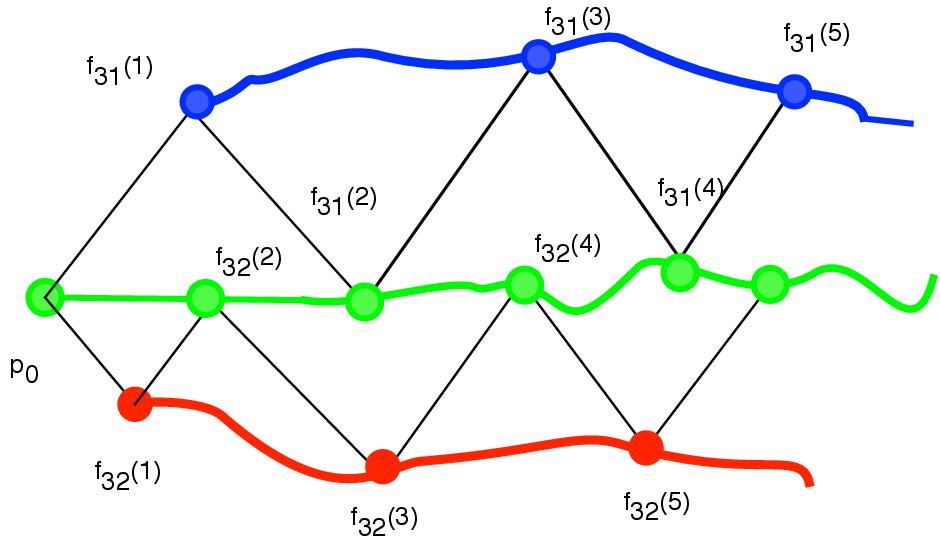} 
\caption{ Illustrated is the trajectory of charge $3$ (green) and
trajectories of charges $1$ (blue) and $2$ (red), pictorially displayed
along the time direction, even though trajectories $1$ and $2$ are bounded.
The acceleration is discontinuous at point $p_{0}$ along trajectory $3$,
which is connected by the advanced interaction to point $f_{31}(1)$ along
trajectory $1$, producing a forward sewing chain of breaking points, i.e.,
points $f_{31}(2)$, $f_{31}(3)$, $f_{31}(4)$ and $f_{31}(5)$. Illustrated is
also a sewing chain of discontinuities starting from point $p_{0}$ and
proceeding in lightcone to trajectory $2$, i.e., points $f_{32}(1)$, $%
f_{32}(2)$, $f_{32}(3)$, $f_{32}(4)$ and $f_{32}(5)$. }
\label{Fig1}
\end{figure}

\bigskip

\bigskip

The acceleration discontinuities along each bound trajectory oscillate
periodically, so that an horizontally traveling third charge interacts with
the discontinuity points along each bound orbit simultaneously (by
symmetry), so that kicks received from each charge can cancel each other by
synchronization. Therefore, the forward scattering direction is a solution
by symmetry. Placing the asymptotically scattered
constant-velocity in another direction on the right-hand-side of Fig. 2 \ is
tricky because the scattered electron interacts with the discontinuities
along each bound trajectory at \emph{different times}. As discussed in
Section \ref{Section III}, the scattered  velocity
must jump to leave the Einstein-local momentum (\ref{continuity}) continuous.
The options for scattering are the resonant Bragg directions illustrated in Fig. 2, where
charge $3$ interacts \emph{simultaneously} with the velocity discontinuities
along both bound trajectories, so that condition (\ref{continuity}) is
automatically satisfied with a continuous constant velocity for charge $3$.
This simple microscopic model is in qualitative agreement with the
experimental results on quantum interference. We stress that the sewing
chain oscillations illustrated in Fig. 1 are natural for our
advanced/retarded equations, \emph{unlike} the case of a model using an
instantaneous ODE. For example, the popular model using an ODE with
instantaneous short-ranged Coulomb interactions is excluded, an in that
respect even an ODE with long-range instantaneous interactions does not
display oscillatory sewing chains.

\begin{figure}[h]
\centering
\includegraphics[scale=0.37]{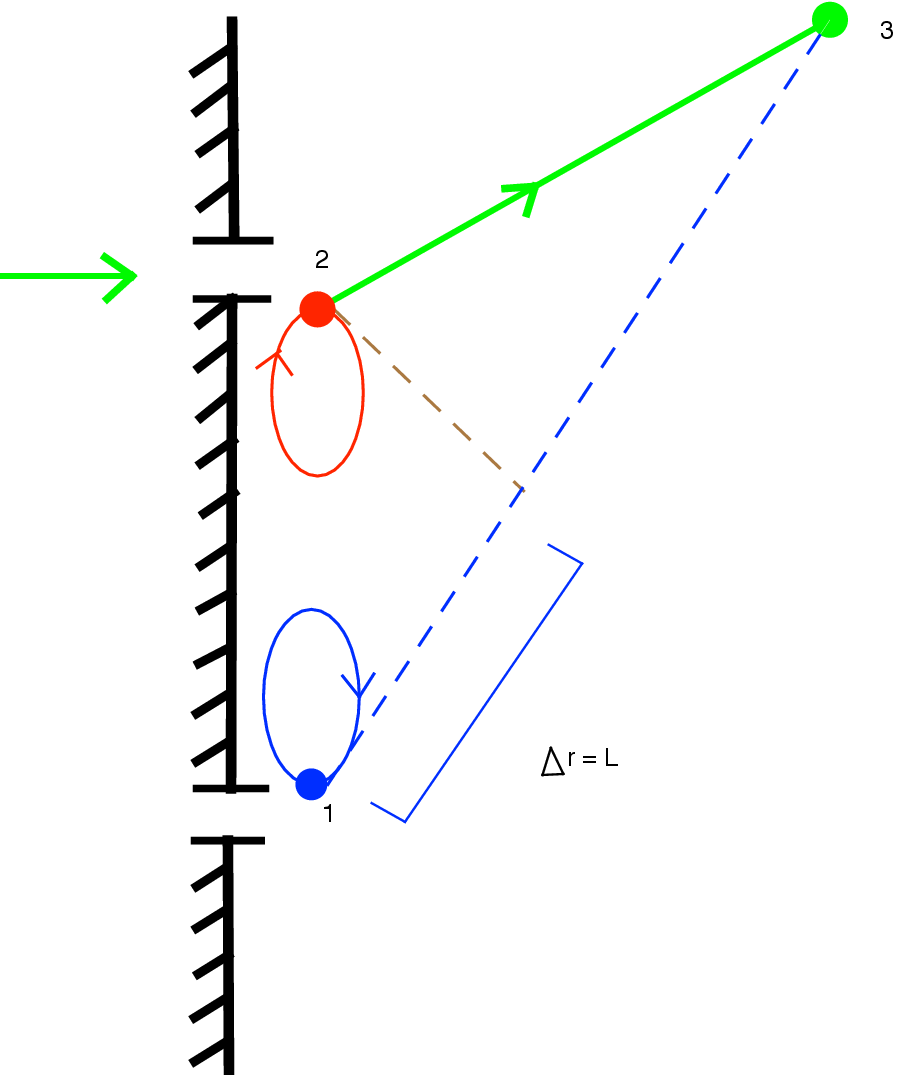} 
\caption{ Illustrated are the bound trajectories of electrons $1$ (solid
blue) and $2$ (solid red) at each material slit and the
piecewise-straight-line trajectory of the scattered electron $3$ (solid
green). Illustrated is the distance from particle $3$ to particle $2$
(green) and the distance from particle $3$ to particle $1$ (blue). The
different distances from electron $3$ to the bound trajectories define a
different delay for each interaction, which resonates when this difference
is equal to the period $L$. }
\label{Fig2}
\end{figure}
Our simplified model explains the qualitative properties of quantum
double-slit scattering; namely (i) Existence of favored lateral directions
of scattering controlled by Bragg interference, which is due to the
state-dependent delay, (ii) Existence of a forward direction of scattering
by the double-slit system, and (iii) Scattering with one slit closed at a
time defines only a forward scattering peak while scattering with both slits
open displays interference. The interference is due to the long-range nature
of the forces and is present in any model with interaction-at-a-distance.
The specific form of our interference turned out to be wavelike \emph{because%
} we used the electrodynamics of point-charges, which inherited
state-dependent-delay from Maxwell's wave theory. Other models with
long-range interactions should display interference as well, but perhaps not
precisely wavelike. In the next section we discuss a far-field
model with piecewise-constant-velocities that predicts the inverse
dependence of the De Broglie wavelength on the incoming particle velocity
and yields a value in the correct order of magnitude.

\bigskip

\section{Piecewise-Defined-Solutions and Far-Field-Model}

\label{Section III}

In the Wheeler-Feynman electrodynamics one does not solve differential
equations for the electromagnetic fields, but rather the trajectories are
the critical points (minimizers) of a variational method with mixed-type
boundaries \cite{JMP2009}. The electromagnetic fields are simply the
coupling terms to the other particle's trajectories as defined by the Euler-Lagrange
equations of the variational method. Rigorously speaking, these 
Euler-Lagrange-coupling-fields are defined only \emph{on} the trajectories,
even though given by the exact same usual formulas of Maxwell's
electrodynamics \cite{JMP2009}. \ Our model uses the equations of motion of
point charges keeping only the far-field interactions. We disregard the
short-range Coulomb interactions that are approximately cancelled by charge
neutrality and ignore the short-range Biot-Savart interactions that are
small at low velocities. In a unit system where $c=e=1$, the low-velocity
Wheeler-Feynman equations of motion can be expressed in the familiar form 
\cite{JMP2009} 
\begin{equation}
m_{k}\frac{d^{2}}{dt^{2}}\mathbf{x}_{k}(t)=-\mathbf{E}(\boldsymbol{\mathbf{x}%
}_{k}\boldsymbol{\mathbf{,}}t)-\mathbf{v}_{k}\times \mathbf{B}(\boldsymbol{%
\mathbf{x}}_{k}\boldsymbol{\mathbf{,}}t),  \label{neumotion}
\end{equation}%
where $\mathbf{v}_{k}\equiv d\mathbf{x}_{k}(t)/dt$. The electric
far-field-coupling-term for each point-charge in the Wheeler-Feynman
electrodynamics is given by a semi-sum of advanced and retarded fields \cite%
{Fey-Whe},

\begin{equation}
\boldsymbol{E}(\boldsymbol{\mathbf{x,}}t)=\frac{1}{2}\mathbf{E}^{+}(%
\boldsymbol{\mathbf{x,}}t)+\frac{1}{2}\mathbf{E}^{-}(\boldsymbol{\mathbf{x,}}%
t),
\end{equation}%
while the far-magnetic field is given by 
\begin{equation}
\boldsymbol{B}(\boldsymbol{\mathbf{x,}}t)=\frac{1}{2}\boldsymbol{n}%
_{+}\times \mathbf{E}^{+}(\boldsymbol{\mathbf{x,}}t)-\frac{1}{2}\boldsymbol{n%
}_{-}\times \mathbf{E}^{-}(\boldsymbol{\mathbf{x,}}t),
\end{equation}%
where unit vectors $n_{k}^{\pm }\equiv (\mathbf{x}-\mathbf{x}_{2}(t_{k}^{\pm
})/|\mathbf{x}-\mathbf{x}_{k}(t_{k}^{\pm })\mathbf{|}$ point respectively
from the charge's advanced/retarded position $\mathbf{x}_{k}(t_{k})$ to
point $\boldsymbol{\mathbf{x}}$ at time $t$ ( see Chapter 14 of \cite%
{Jackson}). The far-electric and far-magnetic field of each charge $k$\ are
defined piecewise by the Li\'{e}nard-Wiechert formulas \cite{Jackson}

\begin{equation}
\mathbf{E}_{k}^{\pm }(\boldsymbol{\mathbf{x,}}t)=\frac{-\mathbf{n}_{k}^{\pm }%
}{r_{k}^{\pm }}\times \frac{\lbrack (\mathbf{n}_{k}^{\pm }\pm \mathbf{%
\mathbf{v}}_{k}^{\pm }\mathbf{)}\times \mathbf{a}_{k}^{\pm }]}{(1\pm \mathbf{%
n}_{k}^{\pm }\cdot \mathbf{v}_{k}^{\pm })^{3}},  \label{far-electric1}
\end{equation}%
and%
\begin{equation}
\mathbf{B}_{k}^{\pm }(\boldsymbol{\mathbf{x,}}t)=\mp \frac{\mathbf{n}%
_{k}^{\pm }}{r_{k}^{\pm }}\times \lbrack \frac{\mathbf{a}_{k}^{\pm }}{(1\pm 
\mathbf{n}_{k}^{\pm }\cdot \mathbf{v}_{k}^{\pm })^{2}}\mp \frac{(\mathbf{n}%
_{k}^{\pm }\cdot \mathbf{a}_{k}^{\pm })\mathbf{\mathbf{v}}_{k}^{\pm }}{(1\pm 
\mathbf{n}_{k}^{\pm }\cdot \mathbf{v}_{k})^{3}}],  \label{far-magnetic1}
\end{equation}%
where the advanced/retarded times are functions of $t$ and $\mathbf{x}$
given by 
\begin{equation}
t_{k}^{\pm }=t\pm |\mathbf{x}_{k}(t_{k}^{\pm })-\mathbf{x|},
\label{retard_time}
\end{equation}%
and the distances in lightcone are functions of $t$ and $\mathbf{x}$ defined
as 
\begin{equation}
r_{k}^{\pm }=|\mathbf{x}_{k}(t_{k}^{\pm })-\mathbf{x|},
\label{lightconedist}
\end{equation}%
where $\ t_{k}$ is defined by Eq. (\ref{retard_time}). In Eqs. (\ref%
{far-electric1}) and (\ref{far-magnetic1}) $\mathbf{v}_{k}^{\pm }\equiv d%
\mathbf{x}_{k}/dt_{k}|_{t_{k}^{\pm }}$ and $\mathbf{a}_{k}^{\pm }$ $\equiv
d^{2}\mathbf{x}_{k}/dt_{k}^{2}|_{t_{k}^{\pm }}$ are respectively the
charge's velocity and acceleration at the advanced/retarded time $t_{k}^{\pm
}$. It is possible to show that the derivative of the advanced/retarded time
respect to $t$ is given by%
\begin{equation}
\frac{dt_{k}}{dt}=\frac{1}{(1\pm \mathbf{n}_{k}^{\pm }\cdot \mathbf{v}%
_{k}^{\pm })},  \label{time_derivative}
\end{equation}%
using Eq. (\ref{retard_time}) and the chain rule. Using Eqs. (\ref%
{far-magnetic1}) and (\ref{retard_time}) with the chain rule we can re-write
the far-magnetic field as 
\begin{equation}
\mathbf{B}_{k}^{\pm }(\boldsymbol{\mathbf{x,}}t)=\mp \frac{\mathbf{n}%
_{k}^{\pm }}{r_{k}^{\pm }}\times \frac{\partial^{2}}{\partial t^{2}}\mathbf{x}%
_{k}(t_{k}^{\pm }),  \label{far-simple}
\end{equation}%
where the partial derivative indicates that Eq. (\ref{retard_time}) is to be used with $\mathbf{x}$ as a fixed argument. 
Finally, the far-electric field is given by%
\begin{equation}
\mathbf{E}_{k}^{\pm }(\boldsymbol{\mathbf{x,}}t)=\pm \mathbf{n}_{k}^{\pm
}\times \mathbf{B}_{k}^{\pm }(\boldsymbol{\mathbf{x,}}t) .
\label{far-elec_simple}
\end{equation}

The far-field at each charge is the semi-sum of the advanced and the
retarded fields (\ref{far-simple}) and (\ref{far-elec_simple}) of the other
two charges. For example, the low-velocity equation of motion of charge $1$,
(Eq. (\ref{neumotion}) with $k=1$), is a simple-looking neutral differential
delay equation 
\begin{eqnarray}
m_{1}\frac{d^{2}}{dt^{2}}\mathbf{x}_{1}(t) &=&\frac{(n_{12}^{+}+\mathbf{v}%
_{1}^{+})}{2r_{12}^{+}}\times \lbrack n_{12}^{+}\times \frac{\partial^{2}}{\partial t^{2}}%
\mathbf{x}_{2}(t_{2}^{+})]  \label{neutral_motion} \\
&&+\frac{(n_{12}^{-}-\mathbf{v}_{1}^{-})}{2r_{12}^{-}}\times \lbrack
n_{12}^{-}\times \frac{\partial^{2}}{\partial t^{2}}\mathbf{x}_{2}(t_{2}^{-})]  \notag \\
&&+\frac{(n_{13}^{+}+\mathbf{v}_{1}^{+})}{2r_{13}^{+}}\times \lbrack
n_{13}^{+}\times \frac{\partial^{2}}{\partial t^{2}}\mathbf{x}_{3}(t_{3}^{+})]  \notag \\
&&+\frac{(n_{13}^{-}-\mathbf{v}_{1}^{-})}{2r_{13}^{-}}\times \lbrack
n_{13}^{-}\times \frac{\partial^{2}}{\partial t^{2}}\mathbf{x}_{3}(t_{3}^{-})],  \notag
\end{eqnarray}%
where the $\mathbf{x}_{k}(t_{k}^{\pm })$ for $k=2,3$ are implicit functions
of time $t$ by Eqs. (\ref{retard_time}) with $\mathbf{x\equiv x}_{1}(t)$.
The equations of motion for the other (negative) charges are obtained by
exchanging indices in Eq. (\ref{neutral_motion}), and Eq. (\ref{motion 1})
is the low-velocity-limit obtained by dropping the velocity terms of the
right-hand-side of Eq. (\ref{neutral_motion}).

\par
The fields of the heavy positive charges are essential to cancel the
short-range Coulomb interaction and might participate to create vanishing far fields \cite{io}. 
In turn, the motion of the bound electrons, henceforth assumed periodic, is essentially responsible to create a vanishing far-field that leaves the scattered electron undisturbed at
large distances. For this, velocity discontinuities of the bound electrons are absolutely needed, as follows;  As shown in \cite{unpublished}, if a bounded motion of two charges produces a net far-field that vanishes almost everywhere, then their trajectories \emph{must} have
discontinuous velocities on sewing chains \cite{unpublished}. This is the physical need for discontinuous velocities. Vice-versa, the scattered electron can go out in a motion
with piecewise-constant-velocity that produces no far-fields (because of Eqs. (\ref%
{far-electric1}) and (\ref{far-magnetic1})), leaving the bound charges
undisturbed.  An 
approximation capturing the essentials of this five-body-dynamics is
by orbits with discontinuous velocities \cite%
{unpublished} of the electromagnetic variational method \cite{JMP2009}. 
 In the following we approximate the trajectory of scattered charge $3$ by two
segments with piecewise-constant-velocity, having a single point where velocity $%
3$ is discontinuous.
\par
  An orbit with a discontinuous velocity can be a
minimizer of the variational method only if some other velocity is
discontinuous in the past/future lightcone, in a way that conserves an
Einstein-local momentum \cite{unpublished}. For these minimizers, the Wheeler-Feynman equations of motion hold \emph{piecewise}, and at velocity discontinuity 
points an additional condition
for an extremum is the continuity of an
Einstein-local-spin-like-four-momentum\cite{unpublished} $(\gamma_i, P_i) $ with
\begin{eqnarray}
\gamma _{i} &\equiv &\frac{m_{i}c}{\sqrt{c^{2}-\mathbf{v}_{i}^{2}}}
\notag \\
&&-\sum\limits_{j\neq i}[\frac{(\frac{e^{2}}{2c})}{r_{ij-}(c-\mathbf{n}%
_{ij-}\cdot \mathbf{v}_{j-})}+\frac{(\frac{e^{2}}{2c})}{r_{ij+}(c+\mathbf{n}_{ij+}\cdot 
\mathbf{v}_{j+})}],  \notag \\
\label{gammas}
\end{eqnarray}
and vector component 
 \begin{eqnarray}
P_{i} &\equiv &\frac{m_{i}c\mathbf{v}_{i}}{\sqrt{c^{2}-\mathbf{v}_{i}^{2}}} \notag
 \\
&&-\sum\limits_{j\neq i}[\frac{(\frac{e^2}{2c}) \mathbf{v}_{j-}}{r_{ij-}(c-%
\mathbf{n}_{ij-}\cdot \mathbf{v}_{j-})}+\frac{(\frac{e^2}{2c}) \mathbf{v}_{j+}}{r_{ij+}(c+%
\mathbf{n}_{ij+}\cdot \mathbf{v}_{j+})}],  \notag \\
\label{continuity}
\end{eqnarray}
which must be \emph{continuous} at each velocity discontinuity point of 
\emph{each} continuous trajectory, despite the discontinuous velocities on both trajectories \cite{unpublished}. Equations (\ref%
{continuity}) and (\ref{gammas}) are written in a unit system where the
speed of light is $c$ and the electronic charge is $e$, in order to make dimensional 
dependences explicit. In the following we use discontinuous trajectories as
an intuitive approximation tool to estimate the dynamics without having to
solve the detailed state-dependent neutral differential delay equations.

In order to introduce a single discontinuity for the scattered velocity, our
simplification further assumes that at its velocity-discontinuity-point
charge $3$ sees the single discontinuity of velocity $1$ at a distance $L$ 
\emph{and} the discontinuity of velocity $2$ also in lightcone at a larger
distance $nL$. The velocity discontinuities along the bound orbits must
either see each other in lightcone (which requires a resonance), or we must
introduce a larger number of velocity discontinuities. After the collision,
the bound electron collides with the heavy positive charge, avoiding the
destruction of the scattering system by moderating the size of velocity
discontinuities. The remaining excited oscillations have a subsequent
influence on the out-scattered charge. In this situation we can produce the
following estimate for the De Broglie wavelength; We use Eq. (\ref%
{continuity}) to estimate the distance of closest approximation $L$ at the
discontinuity point where charge $3$ collides with the bound charge and its
vertical velocity $3$ jumps from zero to a value of the order of $|\mathbf{v}%
_{3}|$. The vertical component of Eq. (\ref{continuity}) yields the estimate%
\begin{equation}
m_{3}|\mathbf{v}_{3}|\simeq \frac{\frac{e^{2}}{c}|\mathbf{v}_{1-}|}{L(c-%
\mathbf{n}_{31-}\cdot \mathbf{v}_{1-})},  \label{estimateL}
\end{equation}%
where we have assumed $(c-\mathbf{n}_{31-}\cdot \mathbf{v}_{1-})\simeq (c+%
\mathbf{n}_{31+}\cdot \mathbf{v}_{1+})$ and $|\mathbf{v}_{1-}|\simeq |%
\mathbf{v}_{1+}|$. The collision is not necessarily a small perturbation on
the bound electron's dynamics, and the denominator on the right-hand-side of
Eq. (\ref{estimateL}) can be large, as for example when $\mathbf{v}_{1-}$
performs fast solenoidal ping-pong motions \cite{io}. Notice that Eq. (\ref%
{estimateL}) yields the experimentally tested De Broglie length inversely
proportional to the incoming velocity. The bound electron subsequently
collides with the heavy positive charge at the slit end (a proton of mass $%
m_{p}$), sharing the momentum recoil and healing most of its velocity
discontinuity. Otherwise the scattered electron destroys the scattering
system and we are studying a bowling problem. We have at hand a \emph{%
five-body-problem} with \emph{ten} state-dependent delays of neutral type.
Our approximation introduces a single velocity discontinuity along the
scattered trajectory, which must be compensated by a velocity discontinuity
in lightcone along the orbit of the bound electron. Velocity discontinuities
balance in lightcone by producing continuous momenta (Eqs. (\ref{continuity}%
) and (\ref{gammas})) at (the) breaking point(s) along the protonic
trajectory and at (the) breaking point(s) along the trajectory of the bound
electron, i.e.,%
\begin{eqnarray}
\frac{m_{p}c}{\sqrt{c^{2}-\mathbf{v}_{p}^{2}}}-\frac{e^{2}/c}{r(c-\mathbf{n}%
_{pe-}\cdot \mathbf{v}_{1-})} &\simeq &0,  \label{proton} \\
\frac{m_{e}c}{\sqrt{c^{2}-\mathbf{v}_{1-}^{2}}}-\frac{e^{2}/c}{r(c-\mathbf{n}%
_{ep-}\cdot \mathbf{v}_{p})} &\simeq &0,
\end{eqnarray}%
where $r$ is the distance in lightcone at the subsequent collision and $%
\mathbf{v}_{p}$ is the (assumed small) protonic velocity, so that we ignore
denominators depending on $\mathbf{v}_{p}$ in Eqs. (\ref{proton}). \ We
further assume $(c-\mathbf{n}_{pe}\cdot \mathbf{v}_{1-})\simeq (c+\mathbf{n}%
_{pe+}\cdot \mathbf{v}_{1+})\simeq (c-\mathbf{n}_{31-}\cdot \mathbf{v}%
_{1-})\simeq c-|\mathbf{v}_{1-}|$, so that eliminating $r$ from Eqs. (\ref%
{proton}) yields%
\begin{equation}
\frac{c}{(c-\mathbf{n}_{31-}\cdot \mathbf{v}_{1-})}\simeq (\frac{\sqrt{2}%
m_{p}}{m_{e}})^{2/3},  \label{hbar}
\end{equation}%
and the De Broglie length $\lambda _{DB}=L$ defined by Eq. (\ref{estimateL})
is%
\begin{equation}
\lambda _{DB}=(\frac{\sqrt{2}m_{p}}{m_{e}})^{2/3}(\frac{e^{2}}{c})\frac{1}{%
m_{3}|\mathbf{v}_{3}|}.  \label{DeBroglie}
\end{equation}%
Formula (\ref{DeBroglie}) yields a length four times smaller than the
popular double-slit formula, a rough estimate relying on an approximate
solution with discontinuous velocities of variational electrodynamics. In our
 five-body model, the electron interacts with a disturbance it has created, much like
 the experiment of Ref. \cite{Bush}
 
\par
Apart from a
proportionality factor, the dependence $\hslash \propto $ $%
(m_{p}/m_{e})^{2/3}$ is singled out as the only power-law consistent with
the hydrogenoid spectroscopic formula, as follows; The spectroscopic
frequency of the hydrogen spectrum for the $n_{1}\rightarrow n_{2}$
transition is%
\begin{equation}
w_{12}=\frac{m_{e}e^{4}}{2\hbar ^{3}}(\frac{1}{n_{1}^{2}}-\frac{1}{n_{2}^{2}}%
).  \label{balmer}
\end{equation}%
If the protonic mass is multiplied by two, as for a deuterium atom, our
power $2/3$ law predicts a four-times-larger $\hslash ^{3}$, so that the
spectroscopic frequency (\ref{balmer}) scales to another line of the
hydrogen spectrum, i.e., the frequency of the $2n_{1}\rightarrow 2n_{2}$
transition. \ Together with the estimates of the spectroscopic lines \cite%
{io}, estimate (\ref{DeBroglie}) of the De Broglie length suggests that
Wheeler-Feynman electrodynamics generalized to discontinuous velocities can
offer an alternative to complete quantum mechanics. The derivation of a
fundamental formula like Eq. (\ref{DeBroglie}) is foreign to the Machian
mentality of quantum mechanics; Notice that even though Eq. (\ref{DeBroglie}%
) was derived using a hydrogen atom at each slit end, another model using a
heavier atom with $n$ electrons and a proportionally larger nuclear mass $\
nm_{p}$ yields the same estimate because there are $n$ electrons to share
the recoil in Eqs. (\ref{proton}). A model using a large number of atoms
along the inner gate of length $a$ would also lead to the same estimate, so
that formula (\ref{DeBroglie}) is somewhat "universal".

\textbf{\bigskip }

\section{Discussions and conclusion}

\label{Discussions}

The danger of replacing the Wheeler-Feynman equations of motion\cite{Fey-Whe} by a Coulombian ODE, besides lack of Poincar\'{e} invariance, are that these are only part of the conditions for a minimizer of the boundary-value variational method with boundaries in past and \emph{future} \cite{unpublished}. As for the fact that these two-body equations of motion \emph{seem} to reduce ``formally'' to a Coulombian ODE at low velocities and large
separations, the qualitative theory of appendix A teaches us that reduction to an ODE with  $C^{\infty }$ trajectories is not granted with a neutral differential delay equation\cite{unpublished}. 
From the perspective of the variational method with boundaries in past and \emph{future} \cite{JMP2009,unpublished},  the set of boundary conditions leading to a many-body dynamics 
with $C^{\infty }$ trajectories is probably a set of measure zero. As discussed in appendices A and B,
this is certainly not the most general set of initial conditions and it is
unlikely that initial conditions leading to $C^{\infty }$ smooth motions
with no velocity/acceleration discontinuities are ever going to be met in
experiments. \ More likely a larger class of initial conditions may
lead to solutions where discontinuities do not explode asymptotically and
stay bounded by laboratory walls ( i.e., respect the constraints of the
microscopic model). The fact that the electromagnetic variational method has boundaries in past and \emph{future} \cite{JMP2009,unpublished} is a warning for Bell-type proofs not to assume reduction to an ODE, at variance with the two-point-boundary-value problem of Hamilton's principle of instantaneous Newtonian mechanics. Even reduction to piecewise $C^{2 }$ orbits is an extra hypothesis that narrows down the qualitative possibilities of the boundary value variational method of electrodynamics\cite{unpublished}. We hope to enrich the attempts to complete QM and to
prevent/enlighten attempts at impossibility proofs based on
"microscopic modeling with trajectories".
\par
We have started on detailed modeling motivated by the subsequent works \cite%
{Jauho} that tried to explain the unexpected experimental result\cite{Mahalu}%
. In the light of a detailed model with an interaction-at-a-distance rather
than by contact, the contradictions of the billiard-ball-model disappear.
Unfortunately, the subsequent explanations\cite{Jauho} of the experimental
disagreement\cite{Mahalu} seem to suggest that the double-slit experiment
does not have a simple quantitative/qualitative precision, or it might have
provided a testbed for microscopic modeling. Notice that the Wheeler-Feynman
electrodynamics with neutral differential delay equations is an
Einstein-local theory of interactions; The velocity discontinuities along
sewing chains involve only Einstein-local-two-body-events connected by a
lightcone\cite{JMP2009}, much like "Einstein-local-quantum-jumps". The
agreement of our far-field-and-piecewise-constant-velocities-model with the
De Broglie wavelength is a test for the use of minimizers with discontinuous
velocities to approximate the solutions of the electromagnetic variational
method\cite{JMP2009}. In fact, the success of our simple model of Section \ref{Section III}
to complete quantum mechanics suggests that the 
electromagnetic variational method \cite{JMP2009} is a physically sensible 
completion for the electrodynamics of point-charges. 
\par
Several (if not all) of the popular
failures to complete quantum mechanics with electrodynamics stem from the
fact that electrodynamics of point-charges with $C^{\infty}$ trajectories is an incomplete theory \cite{unpublished}. The
popular reverse-analogy with quantum mechanics suggested in the 1930's an
electrodynamics formally analogous to the instantaneous
Hamiltonian dynamics, a straw-man to generate impossibility proofs and
conundrums. The subsequent search for a quantization procedure starting from a Hamiltonian
lead to the famous stall of the Wheeler-Feynman program\cite{Mehra} in the days before the 
no-interaction theorem\cite{Currie}. The no-interaction
theorem \cite{Currie} of 1963 was probably the first serious obstacle for
the reverse-analogy approach in the days before the dynamical systems theory. 
\par
The electromagnetic variational method\cite{JMP2009} 
with boundary data yielding globally $C^{2}$ orbits contains the
reverse-analogy electrodynamics as the poor cousin. 
Ignoring the dynamical systems theory of appendix A,
several \emph{formal} embeddings of electrodynamics into a renormalized
Hamiltonian dynamics used expansions of delayed arguments about non-generic $%
C^{\infty}$ trajectories, expansions that are meaningless along orbits with
discontinuous velocities. Besides expansion of delayed arguments, Dirac's 
derivation of the self-force\cite{Dirac} used Poynting's theorem, which holds only for 
globally $C^{2}$ orbits\cite{unpublished}. The abundance of runaways of the Dirac electrodynamics 
of point charges again suggests that physics lies with the
discontinuous-velocity-solutions of the
boundary-value-variational-electrodynamics\cite{JMP2009}. 
\par

The leading motivation to construct the boundary-value-variational-method\cite%
{JMP2009} was to pave the way to study the electromagnetic-two-body-problem 
\emph{numerically}. Additional tests for the variational electrodynamics
with discontinuous velocities would be a microscopic model for Rutherford
scattering. Also a detailed model for Compton scattering on wax\cite{Dodd}
should include many electrons and a heavy nucleus, as discussed below Eq. (%
\ref{balmer}) of Section \ref{Section III}. Further understanding on this neutrally-delayed
dynamics should come after the development of a robust numerical integrator.
A variational integrator shall be published elsewhere along with some
numerical experiments.

\bigskip

\section{Acknowledgements}

\bigskip

We acknowledge partial support from FAPEP and CNPQ. We thank an anonymous
referee from FAPESP that, in the year 2006, refused an unrelated project of
ours on grounds that some day we might try to explain diffraction using
electrodynamics. The author is solely responsible for opinions or whatever
errors may yet linger in the manuscript. We acknowledge interesting
discussions at an early stage with Karl Hess, Michael Mackey and Paulo
Farinas. We thank Tony Humphries and Savio Rodrigues for a numerical
integrator that first faced us with discontinuous accelerations. We thank
Marcel Novaes, Coraci Malta and Karl Hess for reading the later version of
the manuscript. \bigskip

\section{\protect\bigskip Figure Captions}

\bigskip

Figure 1; Illustrated is the trajectory of charge $3$ (green) and
trajectories of charges $1$ (blue) and $2$ (red), pictorially displayed
along the time direction, even though trajectories $1$ and $2$ are bounded.
The acceleration is discontinuous at point $p_{0}$ along trajectory $3$,
which is connected by the advanced interaction to point $f_{31}(1)$ along
trajectory $1$, producing a forward sewing chain of breaking points, i.e.,
points $f_{31}(2)$, $f_{31}(3)$, $f_{31}(4)$ and $f_{31}(5)$. Illustrated is
also a sewing chain of discontinuities starting from point $p_{0}$ and
proceeding in lightcone to trajectory $2$, i.e., points $f_{32}(1)$, $%
f_{32}(2)$, $f_{32}(3)$, $f_{32}(4)$ and $f_{32}(5)$. \bigskip

Figure 2; Illustrated are the bound trajectories of electrons $1$ (solid
blue) and $2$ (solid red) at each material slit and the
piecewise-straight-line trajectory of the scattered electron $3$ (solid
green). Illustrated is the distance from particle $3$ to particle $2$
(green) and the distance from particle $3$ to particle $1$ (blue). The
different distances from electron $3$ to the bound trajectories define a
different delay for each interaction, which resonates when this difference
is equal to the period $L$.

\bigskip

Figure 3; Illustrated in red is a $C^{1}$ initial condition for a scalar
delay-differential-equation and the orbit (in black) after $t=0$. The
derivative from the left at the initial time $t=0$ is given by the history
and in general differs from the derivative from the right along the
continuous orbit. At $t=1$ the solution is already $C^{1}$ .

\bigskip

Figure 4; Illustrated in red is a $C^{1}$ initial condition for a scalar
neutral-delay-differential-equation and the orbit (in black) after $t=0$.
The derivative from the left at $t=0$ is given by the history and in general
differs from the derivative from the right along the continuous orbit. At $%
t=1$ the solution still has a kink in the derivative, a discontinuity that generically
propagates to the breaking point at $t=2 $ and to all successive breaking points.

\bigskip

\bigskip

\section{\protect\bigskip Appendix A: Delay- and Neutral-Differential Delay
Equations}

\label{appendix A}

Since the electromagnetic equations of motion are state-dependent neutral
differential delay equations \cite{JMP2009}, in the following we discuss
briefly the qualitative properties of delay- and neutral differential delay
equations \cite{BellenZennaro,JackHale,Hans-Otto}. We start with constant
delay to simplify the exposition. A delay equation for a scalar function is
defined as%
\begin{equation}
\frac{dy}{dt}=F(y(t),y(t-1)),  \label{delayeq}
\end{equation}%
where $F(a,b)$ is usually a given $C^{\infty }$ function. The
right-hand-side of Eq. (\ref{delayeq}) depends of the present value $y(t)$
and the past value $y(t-1)$, so that to construct a unique solution to Eq. (%
\ref{delayeq}) starting from $t=0$ one needs to provide $y(t)$ on the entire
segment $t\in (-1,0)$. In Figure 3 a typical $C^{1}$ initial condition is
illustrated in red while the ensuing continuous trajectory is illustrated in
black. Notice in Fig. 3 that the derivative from the left at $t=0$ is given
by the derivative of the arbitrary history (red) while the derivative from
the right at $t=0$ is $F(y(0),y(-1))$. The necessary initial history segment
and a discontinuous derivative at the starting point are not seen in the
simpler integration of ordinary differential equations. For an ODE the
initial value $y(0)$ determines a unique solution up to a maximum time $%
t_{F} $ where $F(y)$ ceases to be Lipshitz-continuous. In solving Eq. (\ref%
{delayeq}) with an arbitrary $C^{1}$ history there is a discontinuous
derivative at the starting point, so that one can not impose Eq. (\ref%
{delayeq}) at $t=0$. Assuming the initial condition is such that a solution
exists, things are no so bad because Eq. (\ref{delayeq}) holds onwards, and
moreover the kink at $t=0$ smooths out, as follows. At $t=1$ when Eq. (\ref%
{delayeq}) first accesses $t=0$ as a past point, the left- and right-
derivatives predicted by Eq. (\ref{delayeq}) are equal precisely because the
trajectory is continuous, so that the solution $y(t)$ is $C^{1}$ at $t=1$.
Successively, when the smoothed point $t=1$ is first accessed by Eq. (\ref%
{delayeq}) at $t=2$ the derivative of Eq. (\ref{delayeq}) exists, so that
the solution became $C^{2}$, and so on, acquiring more derivatives at
successive points where $t$ is a multiple of the delay, henceforth called
"breaking points" \cite{BellenZennaro,JackHale,Hans-Otto}. It is usual in
solving delay equations numerically \cite{BellenZennaro} to restrict to
derivatives from the left and derivatives from the right at breaking points.
Last, for equations with state-dependent delay the breaking points depend on
the solution, instead of being simple equally-spaced multiples of the
constant delay. If the deviating argument is a monotonically increasing
function of time, the above smoothing is also the generic behavior expected
for state-dependent-delay.

\begin{figure}[h]
\centering
\includegraphics[scale=0.57]{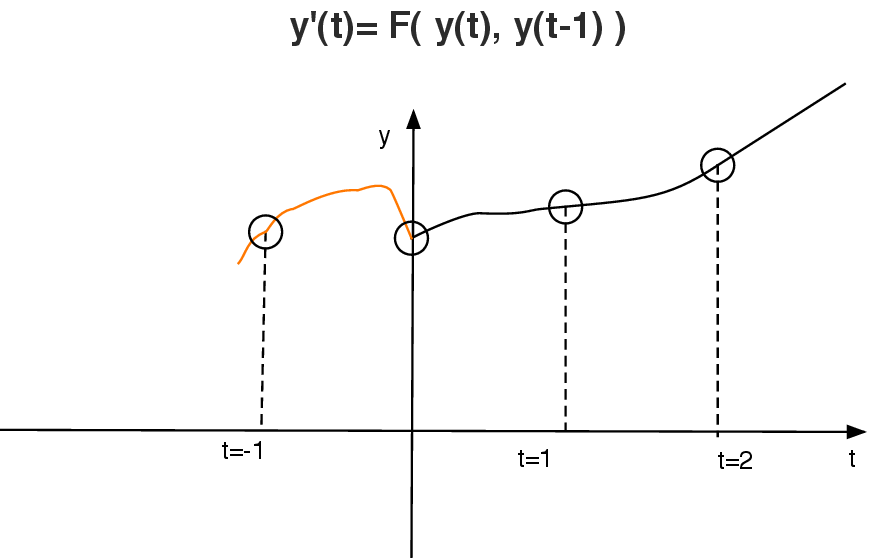} 
\caption{ Illustrated in red is a $C^{1}$ initial condition for a scalar
delay-differential-equation and the orbit (in black) after $t=0$. The
derivative from the left at the initial time $t=0$ is given by the history
and in general differs from the derivative from the right along the
continuous orbit. At $t=1$ the solution is already $C^{1}$ .}
\label{Fig3}
\end{figure}

The integration of a \emph{neutral differential delay equation} for a scalar
function with generic initial data is different, as follows; We start again
with constant delay to simplify the exposition \cite%
{BellenZennaro,JackHale,Hans-Otto}. A neutral differential delay equation
for a scalar function is defined as%
\begin{equation}
\frac{dy}{dt}=G(y(t),y(t-1),\dot{y}(t-1)),  \label{neutraleq}
\end{equation}%
where $G(a,b,c)$ is a given $C^{\infty }$ function. The right-hand-side of
Eq. (\ref{neutraleq}) depends on the present value$\ y(t)$, on the past
value $y(t-1)$ \emph{and} on the past value of the \emph{derivative} $\dot{y}%
(t-1)$. A differential delay equation is said to be a neutral differential
delay equation when the derivative depends on a past value of the
derivative\ itself, as seen in Eq. (\ref{neutraleq}). To construct a unique
solution to Eq. (\ref{neutraleq}) starting from $t=0$ again we need to
provide $y(t)$ for $t\in (-1,0)$ and moreover a $C^{1}$ initial history is
needed because Eq. (\ref{neutraleq}) uses the past derivative. As we already
know, the constructed derivative is discontinuous, as illustrated in Fig. 4
for a typical $C^{1}$ initial history (red). In Fig. 4 the ensuing
continuous trajectory is illustrated in black. Notice that when point $t=0$
is first accessed at $t=1$ along the continuous solution, the derivatives
predicted by Eq. (\ref{neutraleq}) from the left and from right of $t=1$ are
different. We must therefore restrict to left and right derivatives at
breaking points, so that solutions are defined only between breaking points.
At $t=2$ when point $t=1$ is first accessed from the right along the
continuous solution, the derivative discontinuity is still seen by $%
G(y(t),y(t-1),\dot{y}(t-1))$, unlike the case of Eq. (\ref{delayeq}). In
general for a neutral differential delay equation the derivative could stay
discontinuous at all breaking points \cite{BellenZennaro,JackHale,Hans-Otto}%
. Surprisingly, having derivatives undefined on a set of measure zero is
irrelevant for a variational method that involves an integral over the
velocities \cite{JMP2009}, but nevertheless an important qualitative
difference. So much for continuous solutions of one-component delay
equations like Eqs. (\ref{delayeq}) and (\ref{neutraleq}).

\begin{figure}[h]
\centering
\includegraphics[scale=0.57]{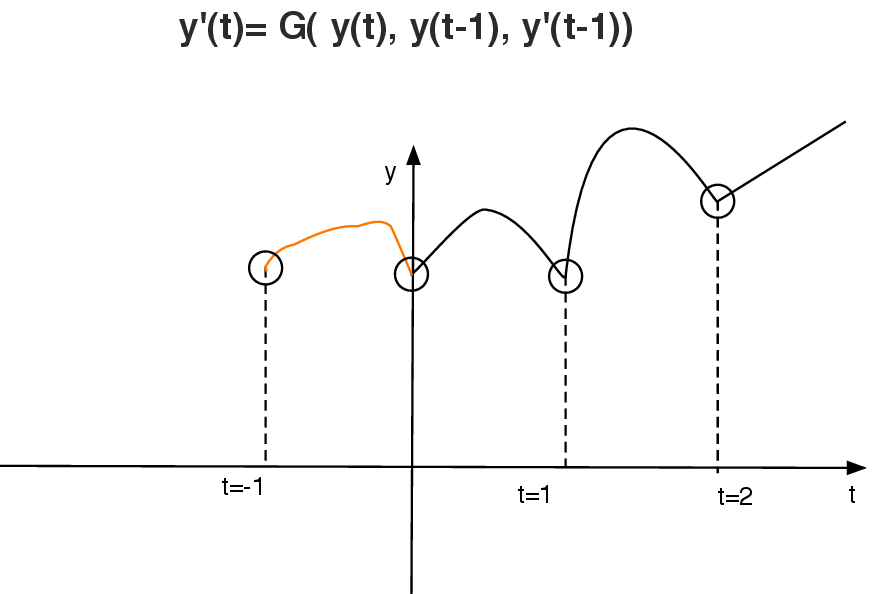} 
\caption{ Illustrated in red is a $C^{1}$ initial condition for a scalar
neutral differential delay equation and the orbit (in black) after $t=0$.
The derivative from the left at $t=0$ is given by the history and in general
differs from the derivative from the right along the continuous orbit. At $%
t=1$ the solution still has a kink in the derivative, which is general does
not vanish and propagates to the next breaking point at $t=2 $ and to all successive breaking points. }
\label{Fig4}
\end{figure}

Again, the difference to neutral differential delay equations with
state-dependent delay is that breaking points depend on the solution,
instead of being simple equally-spaced multiples of the constant delay. In
the general case of neutral differential delay equation, there is the extra
complexity of solution termination \cite{BellenZennaro} when the deviating argument 
are \emph{not} monotonically increasing functions of time. Otherwise continuation is possible 
with discontinuous derivatives. In the case of electrodynamics, deviating arguments 
are always increasing \cite{JMP2009}, so termination does not happen and continuation
is always possible.  In order to make the forward continuation
well-posed, one needs a rule to calculate each velocity discontinuity, like
for example our Einstein-local momentum continuity Eqs. (\ref{continuity})
and (\ref{gammas}) of Section \ref{Section III}. The Wheeler-Feynman equations for the
two-body problem are expressible in the form (\ref{neutraleq}) using a $12$%
-component vector $y(t)$, particle positions taking up $6$ components
while particle velocities take up the other $6$ components. The smoothing properties 
of solutions of many-component differential delay equations
are similar to the neutral differential delay case \cite{ChrisPaul}. For example, 
the $18$-component retarded-only two-body equations of motion of
Dirac's electrodynamics \cite{Dirac} can have solutions along which some
components are continuous (e.g. the positions) while the velocities and accelerations 
are discontinuous at \emph{all} the subsequent breaking points \cite{ChrisPaul}.

\bigskip

\section{ Appendix B: Electron (or neutron) scattered by a crystal}

\label{appendix B}
\bigskip

We give an example that illustrates how a wave theory is not needed
for the explanation of electronic diffraction by crystals \cite{Davisson}, a
simple but not so popular result. Unlike double-slit scattering, the
interference found in crystalline scattering can be explained by a
billiard-ball Hamiltonian model with instantaneous contact forces, even
though certainly not a fundamental explanation. Next and in sake of an
explanation compatible with the theory of relativity, we discuss modeling
crystalline scattering with neutral differential delay equations,
analogously to Section \ref{Section II}. The simple model of Section \ref%
{Section II} uses very little of electrodynamics and is based on a simple
Einstein-local continuity of momentum. Any Einstein-local theory for nuclear
interactions could be used with the same model to explain neutron
diffraction \cite{Shull}. Several early works have explained diffraction by
a crystal using the old quantum mechanics, which is based in Hamiltonian
mechanics \cite{Duane,Compton}.

Our Hamiltonian model assumes that the incoming electron interacts with the
bound electrons of a generic crystal through an instantaneous Hamiltonian
interaction with a periodic potential. We consider a generic Hamiltonian
ordinary differential equation (ODE) with a periodic potential, i.e., 
\begin{equation}
H=\frac{1}{2}|\boldsymbol{p}|^{2}+\epsilon \sum_{_{\mathbf{G}}}V_{\mathbf{G}%
}\exp (i\boldsymbol{G}\cdot \boldsymbol{x}),  \label{defH}
\end{equation}%
where summation is over reciprocal vectors of any generic periodic lattice.
We use complex notation in Eq. (\ref{defH}), with the condition that the
potential is real, i.e., $V_{-\boldsymbol{G}}=V_{\boldsymbol{G}}^{\ast }$,
and $\epsilon $ is included because we assume the potential is small. The
dynamics of Hamiltonian (\ref{defH}) is not integrable in general but it can
nevertheless be simplified with a canonical transformation designed to
remove the potential, i.e., to transform the Hamiltonian into the free
particle Hamiltonian to lowest order in $\epsilon $. The conditions for this
simplification are as follows; We choose a simplifying quasi-identity
canonical transformation \cite{Lichtenberg}, with generating function
depending on the old coordinate $\boldsymbol{x}$ and new momentum $%
\boldsymbol{P}$ 
\begin{equation}
F(\boldsymbol{x},\boldsymbol{P})=\boldsymbol{x}\cdot \boldsymbol{P}+\epsilon
\sum_{_{\mathbf{G}}}F_{\boldsymbol{G}}\exp (i\boldsymbol{G}\cdot \boldsymbol{%
x}),  \label{defF}
\end{equation}%
where the $F_{\boldsymbol{G}}$ are so far arbitrary, respecting only the
reality condition $F_{-\boldsymbol{G}}=F_{\boldsymbol{G}}^{\ast }$. The old
momentum $\boldsymbol{p}$ is related to the new $\boldsymbol{P}$ by%
\begin{equation}
\boldsymbol{p}=\frac{\partial F}{\partial \boldsymbol{x}}=\boldsymbol{P}%
+\epsilon \sum_{_{\mathbf{G}}}i\boldsymbol{G}F_{\boldsymbol{G}}\exp (i%
\boldsymbol{G}\cdot \boldsymbol{x}).  \label{defp}
\end{equation}%
Substitution of $\boldsymbol{p}$ given by Eq. (\ref{defp}) into Eq. (\ref%
{defH}) yields, up to the first order in $\epsilon $, that the new
Hamiltonian is%
\begin{equation}
\tilde{H}=\frac{1}{2}|\boldsymbol{P}|^{2}+\epsilon \sum_{_{\mathbf{G}}}[V_{%
\mathbf{G}}+i(\boldsymbol{G}\cdot \boldsymbol{P})F_{G}]\exp (i\boldsymbol{G}%
\cdot \boldsymbol{x})+O(\epsilon ^{2}).  \label{newH}
\end{equation}%
Hamiltonian $\tilde{H}$ of Eq. (\ref{newH}) is still expressed in terms of
the old coordinate $\boldsymbol{x}$ and the substitution $\boldsymbol{x=X}$
into Eq. (\ref{newH}) is only an $O(\epsilon ^{2})$ mistake. We want to fix
the coefficients $F_{\boldsymbol{G}}$ to vanish the first-order terms of $%
\tilde{H}$ as given by Eq. (\ref{newH}), so that the new momentum $%
\boldsymbol{P}$ is a constant of motion. First let us choose this constant
vector $\boldsymbol{P}$ such that $\boldsymbol{G}\cdot \boldsymbol{P\neq 0}$
for all vectors of the reciprocal lattice, a procedure yielding the $F_{%
\boldsymbol{G}}$ by 
\begin{equation}
F_{G}=\frac{-iV_{\mathbf{G}}}{\boldsymbol{G}\cdot \boldsymbol{P}},
\label{FG}
\end{equation}%
and the new Hamiltonian, Eq. (\ref{newH}), becomes the free-particle
Hamiltonian up to the first order in $\epsilon $, i.e.; 
\begin{equation}
\tilde{H}=\frac{1}{2}|\boldsymbol{P}|^{2}+O(\epsilon ^{2}),  \label{free}
\end{equation}%
having straight-line trajectories up to times of order $1/\epsilon ^{2}$ and
a constant $\mathbf{P}$ as already mentioned. The conclusion is that the
trajectory is an approximate straight-line for most initial conditions,
i.e., if $\boldsymbol{G}\cdot \boldsymbol{P\neq 0}$. These trajectories
account for the forward beam, and constitute the most intense peak of
scattering, in agreement with experiments. On the other hand, if the initial 
$\boldsymbol{P}$ is \emph{perpendicular }to a set of $\boldsymbol{G}^{\prime
}s$ of the reciprocal lattice, it is not possible to remove the
corresponding first-order term of Eq. (\ref{defH}). In this case the
simplified Hamiltonian has a first-order potential to scatter the particle.
In two spatial dimensions, the set of directions satisfying $\boldsymbol{G}%
\cdot \boldsymbol{P=0}$ for a given $\boldsymbol{P}$ are collinear and the
resonant $\boldsymbol{G}^{\prime }s$ are parallel, otherwise the following
simplification is only an approximation; We further assume the $V_{\mathbf{G}%
}$ decrease sufficiently fast with $|\boldsymbol{G|}$, so that we can take
the $\boldsymbol{G}_{o}$ with the least modulus as the essential \emph{%
unremovable} term of the potential, which yields 
\begin{equation}
\tilde{H}=\frac{1}{2}|\boldsymbol{P}|^{2}+2\epsilon |V_{\mathbf{G}_{o}}|\cos
(\boldsymbol{G}_{o}\cdot \boldsymbol{x}),  \label{normal}
\end{equation}%
a simple pendulum Hamiltonian. The higher-order terms produce crossings of
the hyperbolic manifolds of the pendulum Hamiltonian in the usual way \cite%
{Chirikov, Lichtenberg}. The equation of motion for the $\boldsymbol{P}$ of
the normalized Hamiltonian (\ref{normal}) is%
\begin{equation}
\frac{d\boldsymbol{P}}{dt}=[2\epsilon |V_{\mathbf{G}_{o}}|\sin (\boldsymbol{G%
}_{o}\cdot \boldsymbol{x})]\boldsymbol{G}_{o},  \label{eqmotion}
\end{equation}%
from which it follows that the momentum kick is%
\begin{equation}
\Delta \boldsymbol{P=G}_{o}\int\limits_{-\infty }^{\infty }2\epsilon |V_{%
\mathbf{G}_{o}}|\sin (\boldsymbol{G}_{o}\cdot \boldsymbol{x})dt,
\label{kick}
\end{equation}%
a vector parallel to the reciprocal-lattice vector $\boldsymbol{G}_{o}$. An
upper bound for the size of the momentum-kick is the maximum momentum
excursion along the separatrix \cite{Chirikov, Lichtenberg}

\begin{equation}
|\Delta \boldsymbol{P|=}|\boldsymbol{P}_{\max }|=\sqrt{4\epsilon |V_{\mathbf{%
G}_{o}}|},  \label{excursion}
\end{equation}%
so that we can re-write Eq. (\ref{kick}) in the form

\bigskip 
\begin{equation}
\Delta \boldsymbol{P\approx (}\frac{\sqrt{4\epsilon |V_{\mathbf{G}_{o}}|}}{|%
\boldsymbol{G}_{o}|})\boldsymbol{G}_{o},  \label{estimate}
\end{equation}%
The interested reader should read the literature \cite{Chirikov, Lichtenberg}
on chaotic scattering with a Hamiltonian ODE. The resulting momentum kicks
are sensitively dependent on initial conditions as regards magnitude but
come in a prescribed direction, i.e., along a vector of the reciprocal
lattice. In conclusion; along these resonant initial conditions, the
particle suffers stochastic momentum-kicks in the direction of a single
reciprocal-lattice vector \cite{Lichtenberg} as defined by Eq. (\ref{kick}).
This is precisely the von Laue condition \cite{Aschcroft} $\Delta 
\boldsymbol{P=\hslash G}$, in full qualitative agreement with the Bragg--von
Laue wavelike scattering of X-rays by crystals \cite{Aschcroft}. Our simple
model lacks an explicit quantitative size of the angular-momentum kicks, and
to calculate the integral (\ref{kick}) we need the trajectories and the
next-order terms that perturb the simple integrable Hamiltonian (\ref{normal}%
) to a chaotic dynamics. We could adjust $V_{\mathbf{G}_{o}}$ to obtain an
experimental angular-momentum-kick of the mechanism to agree with Eq. (\ref%
{estimate}), i.e., $l_{o}=\sqrt{4\epsilon |V_{\mathbf{G}_{o}}/|\boldsymbol{G}%
_{o}|}$, an adjusting beyond our simplified modeling.

We discussed the above Hamiltonian model as a surprise not often stressed,
but even though in qualitative agreement with experiments, modeling with an
instantaneous Hamiltonian ODE is hindered \emph{in principle} by the
non-interaction theorem \cite{Currie}; The non-interaction theorem \cite%
{Currie} states that the only two-body motion describable by a Hamiltonian
ODE with manifest covariance by the Poincar\'{e} group is the free-particle
motion \cite{Currie}. Essentially, Hamiltonian dynamics is instantaneous,
and relativity introduces the finite speed of light, which demands
delay-equations of motion with solutions defined \emph{piecewise}, as
discussed in appendix A. The no-interaction theorem is not a
surprise; delay equations need a history, and only free-moving particles
have a straight-line-past-history that can be reconstructed from point-like
ODE data. Otherwise, for generic initial data (or boundary data) and
nontrivial interactions, we can not expect an ODE to guess an
infinite-dimensional history. Since free particles move with constant
velocity until a collision, the non-interaction theorem suggests that \emph{%
if} a Hamiltonian description is of any use, the physical problem at hand
must have particles moving with piecewise-constant-velocities, like the
scattered charge in our model of Section \ref{Section III}.

In sake of a model compatible with the theory of relativity, in the
following we discuss scattering by a crystal using neutral differential
delay equations and the same model of Section \ref{Section II} (essentially
a non-Hamiltonian business). Our model places one bound electron at each
site of an infinite crystal. The explanation follows in the same way of the
double-slit modeling of Section \ref{Section II}; Again we assume the
incoming charge superimposes a resonant perturbation of period $L$ on the
bound orbits at each crystalline site, in the same way explained in Sections %
\ref{Section II} and \ref{Section III}. The perturbations are again transferred to
the heavy on-site charges to avoid a bowling effect and to heal most of the
velocity discontinuities. There remains a small oscillation, so that the
out-scattered electron suffers delayed periodic kicks from these
synchronized oscillations excited at the crystalline sites. Let $\hat{u}$ be
the unit vector along the scattered velocity and $\Delta \hat{u}$ the change
of this unitary direction upon scattering. The advanced/retarded delay time $%
\pm \Delta t_{j}$ of interaction with the different crystalline sites due to
the site-dependent $(\Delta r_{j})$ distances of retardation is \cite%
{Aschcroft} 
\begin{equation}
\Delta t_{j}=\Delta \hat{u}\cdot \Delta r_{j},  \label{deltaT}
\end{equation}%
in a unit system where the speed of light is $c=1$. The favored directions
of Bragg scattering are those along which $\Delta t_{j}=nL$ \ for an integer 
$n.$ Using Eq. (\ref{deltaT}) and the usual properties of reciprocal lattice
vectors \cite{Aschcroft} that

\begin{equation*}
\boldsymbol{G}\cdot \Delta r_{j}=2\pi n,
\end{equation*}%
we find that the favored (resonant) $\Delta u$ must be along a reciprocal
lattice vector $\boldsymbol{G}$, i.e., 
\begin{equation}
\Delta u=\frac{L|u|}{2\pi }\boldsymbol{G}.  \label{VonLaue}
\end{equation}%
We see that again a simple Einstein-local microscopic model comes in full
qualitative agreement with the Bragg--von Laue wavelike scattering by
crystals \cite{Aschcroft}. The estimate for the De Broglie length $L$ is the
same given in Section \ref{Section III}.

\end{document}